\documentclass{Rinton-P9x6}

\newcommand{\tr}{{\rm tr }}

\begin{document}

\title{The Quantumness of a Set of Quantum States}

\author{Christopher A. Fuchs}

\address{
Bell Labs, Lucent Technologies, Murray Hill, New Jersey 07974, USA
\\
Ctr.\ for Quantum Communication, Tamagawa University, Tokyo
194-8610, Japan }

\author{Masahide Sasaki}

\address{Communications Research Laboratory, Koganei, Tokyo 184-8795, Japan
\\
CREST, Japan Science \& Technology Corp., Shibuya, Tokyo 150-0002,
Japan}


\maketitle

\abstracts{We propose to quantify how ``quantum'' a set of quantum
states is.  The quantumness of a set is the worst-case difficulty
of transmitting the states through a classical communication
channel. Potential applications of this measure arise in quantum
cryptography, where one might like to use an alphabet of states
most sensitive to quantum eavesdropping, and in lab demonstrations
of quantum teleportation, where it is necessary to check that
entanglement has indeed been used.}

\section{Introduction}

How quantum can a single quantum state be?  Does this question make
sense?  One gets the impression it does with even a small perusal of
the quantum-optics literature, where coherent states are often called
``classical'' states of light. Despite the nomenclature, we suggest
there is no robust notion of the classicality of a {\it single\/}
quantum state. Consider any two coherent states $|\alpha\rangle$ and
$|\beta\rangle$.  The inner product of these states is nonzero.
Thus, if a single mode is prepared secretly in one of these states,
there is no automatic device that can amplify the signal reliably
into a two-mode state $|\gamma\rangle|\gamma\rangle$, where
$\gamma=\alpha,\beta$ depending upon the input. Nonorthogonal states
cannot be cloned,\cite{Yuen86} and this holds whether such states are
called ``classical'' or not.

A notion of the {\it quantumness of states\/} can thus only be
attached to a {\it set\/} of states.  The members of a set of
states can be more or less quantum with respect to each other, but
there is no good sense in which each one alone is intrinsically
quantum or not. A set of two nonorthogonal states $|\psi_0\rangle$
and $|\psi_1\rangle$, with $x=|\langle\psi_0|\psi_1\rangle|$,
provides a good example.\cite{Fuchs00}  To set the stage, let us
work within the metaphor of no-cloning.  There, we might take the
degree of ``clonability''\cite{Buzek96} as measure of quantumness
of the two states.

A cloning attempt is a unitary operation $U$ that gives
$|\psi_i\rangle|0\rangle \longrightarrow |\Psi_i\rangle$, $i=0,1$,
where $|\Psi_i\rangle$ is a state whose partial trace over either
subsystem gives identical density operators.  An optimal cloning
attempt is one that maximizes the fidelity between the output and the
wished-for target state
$|\psi_i,\psi_i\rangle\equiv|\psi_i\rangle|\psi_i\rangle$, i.e.,
maximizes $ F_{\mbox{\scriptsize
try}}=\frac{1}{2}|\langle\Psi_0|\psi_0,\psi_0\rangle|^2 +
\frac{1}{2}|\langle\Psi_1|\psi_1,\psi_1\rangle|^2\;.
\label{Artur}
$
It can be shown that\cite{Bruss98}
\be
F_{\mbox{\scriptsize clone}}\; =\; \max_U\; F_{\mbox{\scriptsize
try}} = \; \frac{1}{2}\Big(1+x^3+(1-x^2)\sqrt{1+x^2}\Big)\;.
\label{Christopher}
\ee
Viewing Eq.~(\ref{Christopher}) as a measure of the quantumness of
two states---i.e., the smaller $F_{\mbox{\scriptsize clone}}$, the
more quantum the set of states---one finds that two states are the
{\it most\/} quantum with respect to each other when $x=1/\sqrt{3}$.

Eq.~(\ref{Christopher}), though we will not ultimately adopt it,
exhibits some of the main features a measure of quantumness ought
to have. In particular, two states are the most classical with
respect to each other when they are either orthogonal or
identical. Moreover, the set is most quantum when the states are
somewhere in between, in this case when they are $54.7^\circ$
apart. This point draws the most important contrast between the
notions of quantumness and
distinguishability.\cite{Helstrom76,Fuchs96a} As an example, in
communication theory it is important to understand the best
probability with which a signal can be guessed correctly after a
quantum measurement has been performed on its carrier. For the
case at hand, the measure of optimal distinguishability is then
given by $ P_s =
\frac{1}{2}\left(1+\sqrt{1-x^2}\right)$.\cite{Helstrom76} This
quantity is monotone in the parameter $x$. No measure of
quantumness should have this character.  Instead, quantumness
should capture how difficult it is to make a copy of the quantum
state after some of the information about its identity has been
deposited in another system.

The optimal cloning fidelity in Eq.~(\ref{Christopher}) is not
completely satisfactory for our purposes, though. One reason is
that the idea of cloning does not lead uniquely to
Eq.~(\ref{Artur}). Under minor modifications of the clonability
criterion, the particular $x$ for which two states are the most
quantum with respect to each other changes drastically.  For
instance, by another measure in Ref.\cite{Bruss98}, two states are
the most quantum with respect to each other when they are
$60^\circ$ apart. Furthermore, by neither of these measures do we
see a potentially desirable connection between quantumness and
{\it angle\/} in Hilbert space. If there is a connection, one
might expect the two states to be the most quantum with respect to
each other when they are $45^\circ$ apart---giving rise to the
pleasing slogan: ``Two states are the most classical when they are
$0^\circ$ and $90^\circ$ apart. They are most quantum when they
are halfway in between.''

For these reasons, we adopt a metaphor more akin to eavesdropping
in quantum cryptography.  For any set of pure states ${\cal
S}=\{\Pi_i=|\psi_i\rangle\langle\psi_i|\}$, let us act as if there
is a source emitting systems with states drawn according to a
probability distribution $\pi_i$. (This distribution is an
artifice; ultimately it will be discarded after an optimization.)
The systems are then passed to an eavesdropper who is required to
measure them one by one and thereafter fully discard the
originals. To make sure the latter process is enforced, we might
imagine that the eavesdropper really takes the form of two people,
perhaps Eve and Yves, separated by a classical channel.  Eve may
perform any quantum measurement imaginable, but then Yves will
have access to nothing beyond the classical information obtained
to attempt to reproduce the original state.

The question is, how intact can the states remain in this process?
To gauge the intactness, we take the average fidelity between the
initial and final states. Operationally this corresponds to Yves
handing his newly prepared quantum system back to the source. The
preparer checks to see whether the system has kept its identity;
the probability it passes the test is the fidelity.

Considering the best measurement and resynthesis strategies Eve and
Yves can perform gets us most of the way toward a notion of
quantumness.  The final ingredient is to imagine that the source
makes this task as hard as possible. Conceptually, we do this by
adjusting the probabilities $\pi_i$ so that the maximum average
fidelity is as small as it can be.  The resulting fidelity is what
we take to be the quantumness of $\cal S$.  The intuition behind
this definition is simple. It captures in a clear-cut way how
difficult the eavesdropper's task can be made for reconstructing the
set of states.  And it does this disregarding the more subtle task
of quantifying how much information Eve learns about the state's
identity in the process. In a way, it captures the raw sensitivity
to eavesdropping that can be imparted to the states in $\cal S$.

The problem promoted here has its roots in the ``state estimation''
scenario studied in great detail in the recent
literature.\cite{Massar95}  The main differences are that we have
relaxed the condition that the states in $\cal S$ be associated with
a uniform distribution on Hilbert space, and we have added an extra
optimization over the probability distribution $\pi_i$. Moreover,
the traditional use of ``state estimation'' has been for purposes of
defining a notion of distinguishability for quantum
states.\cite{Helstrom76,Chefles00} As explained above, this is
exactly what we are {\it not\/} trying to get at with a notion of
quantumness.\cite{Barnett01}

\section{Building an Expression for Quantumness}

Imagine $\cal S$ equipped with a probability distribution $\pi_i$.
Such a set of states along with a set of assigned probabilities, we
call an {\it ensemble\/} $\cal P$. (There are no restrictions on the
number of elements in $\cal S$, nor on $d$, the dimension of Hilbert
space for which $|\psi_i\rangle\in{\cal H}_d$.) Eve performs a single
quantum measurement, i.e., some POVM ${\cal E}=\{E_b\}$, on the
signal.

Yves makes use of the information Eve obtains---some explicit index
$b$---by preparing his system in a quantum state $\sigma_b$. Since
the preparation is based solely on classical information, there need
be no restrictions on the mapping ${\cal M}:b\rightarrow\sigma_b$.
For instance, it need not be completely positive, etc. Moreover, the
$\sigma_b$ may be mixed states. This corresponds to the possibility
of a randomized output strategy on the part of Yves. The conjunction
of $\cal E$ and $\cal M$ constitutes a protocol for the
eavesdropping pair.

Supposing the source emits the state $\Pi_i$, and Eve obtains the
outcome $b$ for her measurement, then the fidelity Yves achieves is
$F_{b,i} = \langle\psi_i|\sigma_b|\psi_i\rangle$. However there is
no predictability of Eve's measurement outcome beyond what quantum
mechanics allows.  Similarly, the most we can say about the actual
state the source produces is through the probability distribution
$\pi_i$.  Therefore, the average fidelity for the protocol is
\be
F_{\cal P}({\cal E},{\cal M}) = \sum_{b,i} \pi_i \tr(\Pi_i E_b)
\tr(\Pi_i\sigma_b).
\ee
A convenient intermediate quantity comes from optimizing Yves'
strategy alone. For a given $\cal E$, we define the {\it achievable
fidelity\/} with respect to the measurement to be $F_{\cal P}({\cal
E})=\max_{\cal M} F_{\cal P}({\cal E},{\cal M})$. In analogy to the
quantity known as accessible information\cite{Fuchs96a} in the theory
of quantum channel capacities, let us define the {\it accessible
fidelity\/} of the ensemble $\cal P$ to be
\be
F_{\cal P}=\max_{\cal E} F_{\cal P}({\cal E})\;.
\label{Jeffrey}
\ee
Finally the {\it quantumness\/} of the set $\cal S$ is defined by
\be
Q({\cal S}) = \min_{\{\pi_i\}} F_{\cal P}\;.
\label{Eugene}
\ee

\section{The Accessible Fidelity} \label{AccFid}

It is easy to derive an exact expression for the achievable fidelity
$F_{\cal P}({\cal E})$ for any given $\cal E$. First note that
$F_{\cal P}({\cal E},{\cal M})$ is linear in the $\sigma_b$. Thus,
in any decomposition of $\sigma_b$ into a mixture of pure states, we
might as well delete $\sigma_b$ and replace it with the most
advantageous element in the decomposition. Therefore, it never hurts
to take the $\sigma_b$ to be pure states,
$\sigma_b=|\phi_b\rangle\langle\phi_b|$.

Rewriting $F_{\cal P}({\cal E},{\cal M})$ under this assumption, we
obtain
\be
F_{\cal P}({\cal E},{\cal M}) = \sum_b
\langle\phi_b|M_b|\phi_b\rangle\;,
\label{Tal}
\ee
where $M_b = \sum_i \pi_i \Pi_i E_b\Pi_i = \sum_i \pi_i \tr(\Pi_i
E_b) \Pi_i$. Now the pure states $|\phi_b\rangle$ are arbitrary.
Thus we can optimize each term in Eq.~(\ref{Tal}) separately.  This
is done by remembering that the largest eigenvalue $\lambda_1(A)$ of
any Hermitian operator $A$ can be characterized by $\lambda_1(A) =
\max\, \langle\alpha|A|\alpha\rangle$. Therefore,
\be
F_{\cal P}({\cal E})=\sum_b\, \lambda_1\Big(\sum_i \pi_i \tr(\Pi_i
E_b) \Pi_i\Big)\;.
\label{William}
\ee

Unfortunately, this is where the easy part of the development ends.
No explicit expression for the accessible fidelity exists in general.
The most one can hope is to understand some of its general
properties, a few explicit examples, and perhaps some useful bounds.

In this regard, the first question to ask is how much can said about
the measurements achieving equality in Eq.~(\ref{Jeffrey}). Are we
even sure that optimal measurements exist? The answer is yes, and the
reason is essentially the same as for the existence of an optimal
measurement for the accessible information.\cite{Holevo73b} Because
$\lambda_1(A+B)\le\lambda_1(A)+\lambda_1(B)$, $F({\cal E})$ is a
convex function over the the set of POVMs.  Since the set of POVMs is
a compact set\cite{Holevo73b} and $\lambda_1(A)$ is a continuous
function, it follows that $F_{\cal P}({\cal E})$ achieves its
supremum on an extreme point of the set. Furthermore, by the
reasoning of Refs.\cite{Fujiwara98,Davies78}, we know that for any
extreme point $\cal E$, all the nonvanishing operators $E_b$ within
it must be linearly independent.  Thus, we can restrict the
maximization in Eq.~(\ref{Jeffrey}) to POVMs with no more than $d^2$
outcomes. Finally, because of the subadditivity of $\lambda_1(A)$,
these $d^2$ operators can be chosen to be rank-one.

One might wish for a further refinement in what can be said of
optimal measurements for accessible information.  For instance, that
the number of measurement outcomes not exceed the number of inputs
in analogy to the case of quantum hypothesis
testing.\cite{Helstrom76}  This intuition is captured by asking
rhetorically, what can Eve possibly do better than make her best
guess and pass that information on to Yves?  If such is the case,
though, a proof remains to be seen.  Indeed, because of the
nonlinearity in Eq.~(\ref{Jeffrey}), there may be counterevidence
from the case of accessible information.\cite{Shor00}

This brings up the question of how to draw a comparison between the
success probability in hypothesis testing and the achievable
fidelity for any given measurement.  The usual way of posing the
hypothesis testing problem is to assume a one-to-one correspondence
between inputs $\Pi_i$ and POVM elements $E_i$, each element
signifying the guess one should make.  In that way of writing the
problem, the average success probability $P_s$ takes the form
$P_s=\sum_i \pi_i \tr(\Pi_i E_i)$. Here, however, we cannot make such
a restriction on the number of outcomes. So, we must pose the
hypothesis testing problem in a more general way.

Suppose Eve performs a measurement $\cal E$ and observes outcome $b$
to occur.  This information will cause her to update here
probabilities for the various inputs according to Bayes' rule:
\be
p(i|b) = \frac{p(b,i)}{p(b)} = \frac{\pi_i\tr(\Pi_i E_b)}{\tr(\rho
E_b)}\;,
\label{Howard}
\ee
where $p(b)=\tr(\rho E_b)$ and $\rho = \sum_i \pi_i \Pi_i$. Maximum
likelihood dictates that Eve's success probability will be optimal if
she chooses the value $i$ for which $p(i|b)$ is maximum.  Thus her
average success probability will be
\be
P_s({\cal E}) = \sum_b p(b) \max_i\{p(i|b)\} = \sum_b
\max_i \{\pi_i \tr(\Pi_i E_b)\}\;.
\label{Benjamin}
\ee

Eq.~(\ref{Benjamin}) compares to $F_{\cal P}({\cal E})$ through a
simple inequality.  To see this, note that $F_{\cal P}({\cal
E})=\sum_b p(b) \lambda_1(\rho_b)$, where $\rho_b = \sum_i p(i|b)
\Pi_i$. Suppose $i$=$j$ maximizes $p(i|b)$. Then $ \lambda_1(\rho_b)
= \max\langle\phi|\rho_b|\phi\rangle \ge
\langle\psi_j|\rho_b|\psi_j\rangle = \sum_i p(i|b)
|\langle\psi_j|\psi_i\rangle|^2 \ge p(j|b) + \sum_{i\ne j} p(i|b)
|\langle\psi_j|\psi_i\rangle|^2 \ge \max_i \{p(i|b)\}$. Therefore,
$P_s({\cal E})\le F_{\cal P}({\cal E})$. This inequality is not
tight, however.  For instance, for $\cal P$ describing a uniform
distribution of states on a qubit, $P_s\rightarrow0$, while $F_{\cal
P}=2/3$.\cite{Chefles00}

Tighter bounds, both upper and lower, on Eq.~(\ref{Jeffrey}) would be
useful.  An obvious lower bound comes directly from the convexity of
the $\lambda_1(A)$ function.  Note in particular that $\rho = \sum_b
p(b) \rho_b$. Therefore $F_{\cal P}({\cal E})\ge \lambda_1(\rho)$.
This inequality is generally tighter than the previous one in that it
never falls below $1/d$; moreover, there is a measurement $\cal E$
that achieves it.

A more interesting lower bound comes about by considering the
behavior of $F_{\cal P}({\cal E})$ with respect to the ``square-root
measurement.''\cite{Holevo78} This POVM is constructed from the
ensemble decomposition of $\rho$ by multiplying it from the left and
right by $\rho^{-1/2}$. Inserting this measurement into $F_{\cal
P}({\cal E})$, we obtain
\be
F_{\cal P}\ge F_{\rm PGM}=\sum_i \lambda_1\Big(\sum_j \pi_i\pi_j
\Pi_j \rho^{-1/2}\Pi_i\rho^{-1/2}\Pi_j\Big).
\label{Jeroen}
\ee

\section{Conclusion}

Much more can be said about accessible fidelity and
quantumness,\cite{Fuchs02a} but lack of space prevents us from saying
it here.  We end instead with a question.  One can define the {\it
quantumness of a Hilbert space\/} by
\be
Q_d = \inf_{\cal S} Q({\cal S})\;,
\ee
where the infimum is taken over {\it all\/} sets of states living on
${\cal H}_d$.  One has to wonder whether this quantity might not
indicate a deep defining property for the quantum system itself---its
ultimate ``sensitivity to the touch.''\cite{Fuchs02b}

\section*{Acknowledgments}
Much of this work was carried out at Caltech in Spring 1999 and at
Tamagawa University in Springs 2000/2001.  CAF acknowledges support
of a DuBridge Fellowship and the hospitality of Tamagawa University
during those periods.  MS thanks S. M. Barnett and C. Gilson for
valuable discussions.


\begin{thebibliography}{99}

\bibitem{Yuen86}
H.~P. Yuen, Phys.\ Lett.\ A {\bf 113}, 405 (1986).

\bibitem{Fuchs00}
C.~A. Fuchs, {\tt quant-ph/9810032}.

\bibitem{Buzek96}
V.~Bu\v{z}ek and M.~Hillery, Phys.\ Rev.\ A {\bf 54}, 1844 (1996).

\bibitem{Bruss98}
D.~Bru\ss, {\it et al.},  Phys.\ Rev.\ A {\bf 57}, 2368 (1998).

\bibitem{Helstrom76}
C.~W. Helstrom, {\sl Quantum Detection and Estimation Theory},
(Academic Press, NY, 1976).

\bibitem{Fuchs96a}
C.~A. Fuchs, {\tt quant-ph/9601020}.

\bibitem {Massar95}
S. Massar and S. Popescu, Phys.\ Rev.\ Lett.\ {\bf 74} 1259 (1995).

\bibitem{Chefles00}
A.~Chefles, Contemp.\ Phys.\ {\bf 41}, 401--424 (2000).

\bibitem{Barnett01}
S.~M. Barnett, C.~R. Gilson and M. Sasaki, J. Phys. A {\bf 34}, 6755
(2001).

\bibitem{Holevo73b}
A.~S. Holevo, J.\ Mult.\ Anal.\ {\bf 3}, 337 (1973).

\bibitem{Fujiwara98}
A.~Fujiwara and H.~Nagaoka, IEEE Trans.\ Inf.\ Theory {\bf 44}, 1071
(1998).

\bibitem{Davies78}
E.~B. Davies, IEEE Trans.\ Inf.\ Theory {\bf IT-24}, 596--599.

\bibitem{Shor00}
P.~W. Shor, {\tt quant-ph/0009077}.

\bibitem{Holevo78}
A.~S. Holevo, Theor.\ Prob.\ Appl.\ {\bf 23}, 429 (1978).

\bibitem{Fuchs02a}
C.~A. Fuchs and M.~Sasaki, {\tt quant-ph/0302092}.

\bibitem{Fuchs02b}
C.~A. Fuchs, {\tt quant-ph/0205039}.

\end{thebibliography}
\end{document}